\documentstyle[12pt]{article}

\input{epsf.sty}  
\input{psfig.sty}

\def \Cn{\buildrel \to \over C}
\def \Bbar{{\buildrel \_ \over B}_0}

\def \Vn{\buildrel \to \over V}
\def \Abar{{\buildrel \_ \over A}_0(l)}
\def \s{\scriptstyle}

\def \eqdef{\buildrel \rm def \over =}  
\def \IR{{\rm I\!R}}  
\def \IN{{\rm I\!N}}  
\def \IC{{\rm C}\llap{\vrule height7.1pt width1pt
     depth-.4pt\phantom t}} 

\font\cmss=cmss10
\font\cmsss=cmss10 at 7pt
\def \IZ{\relax\ifmmode\mathchoice
{\hbox{\cmss Z\kern-.4em Z}}{\hbox{\cmss Z\kern-.4em Z}}
{\lower.9pt\hbox{\cmsss Z\kern-.4em Z}}
{\lower1.2pt\hbox{\cmsss Z\kern-.4em Z}}\else{\cmss Z\kern-.4em Z}\fi}

\begin{document}

\title{\vskip -0.4truecm
\vskip 0.5truecm
CAUSAL SET DYNAMICS: A TOY MODEL}

\author{ {\bf A. Criscuolo \thanks{e-mail: criscuol@phys.uu.nl} 
\ {\rm and} H. Waelbroeck
\thanks{e-mail: hwael@nuclecu.unam.mx}
\thanks{On sabbatical leave from the Institute of Nuclear Sciences, UNAM;
Circuito Exterior, C.U.; A. Postal 70-543; Mexico DF 04510}} \\  
Spinoza Instituut \\
Universiteit Utrecht\\
P.O.Box 80.195\\
3508 TD Utrecht, The Netherlands\\}
\date{12th November 1998} 
\maketitle 

\begin{abstract} 
\baselineskip = 18pt
We construct a quantum measure
on the power set of non-cyclic oriented graphs of N points,
drawing inspiration from 1-dimensional directed percolation. 
Quantum interference patterns lead to properties which do not appear 
to have any analogue in classical percolation. Most notably, instead 
of the single phase transition of classical percolation, the quantum 
model displays two distinct crossover points. Between these two points, 
spacetime questions such as: ``does the network percolate?'' have 
no definite or probabilistic answer.
\end{abstract} 

\begin{flushleft}
\end{flushleft}
\pagebreak

\baselineskip = 18pt

\section{Introduction} 

The effort to formulate a discrete theory of Quantum Gravity
has recently recovered some of its appeal, due to results from 
the quantization of general relativity \cite{rovelli}, black hole 
thermodynamics \cite{beckenstein} and string theory \cite{kato}. 
All suggest that the spectrum of excitations of a theory of quantum 
gravity must be discrete. It follows that counting is a natural 
method to define spacetime volume. A minimum framework for a 
discrete model of spacetime geometry brings in two key elements: 
number and order \cite{sorkin90}. Number gives the local conformal 
factor or spacetime volume element. Causal order suffices to 
define light cones and this represents spacetime geometry
up to a conformal factor. If we call ``points'' 
the objects that are being counted, and assume that the causal 
relations between points do not form closed timelike curves, then
they provide us with a structure called a partially ordered set (poset).
 
A poset ${\cal P}$ is a discrete set with a transitive 
acyclic relation, namely a relation $\prec$ such that 
$\forall x, y, z \in {\cal P}$,
$$x \prec y \ \ and \ \ y \prec z \Rightarrow x \prec z, \eqno (1.1)$$
$$x \prec y \ \ and \ \ y \prec x \Rightarrow x = y. \eqno (1.2)$$
For any two points $x, y \in {\cal P}$, the {\it Alexandrov set} 
\cite{alexandrov} or 
{\it interval} $[x, y]$ is defined by
$$[x, y] \eqdef \{ z : x \prec z \prec y \}. \eqno (1.3)$$ 
A partially ordered set is said to be {\it locally finite} if the 
Alexandrov sets are finite; it is then called a {\it Causal Set}. 
The axiom (1.2) ensures that a causal set has no closed timelike curves. 

Several authors have proposed placing the causal set structure at 
the center of a discrete formulation of quantum gravity \cite{bombelli}
\cite{thooft}. Poset-generating models which have been considered 
range from quantum spin network models \cite{markopoulou} to stochastic 
models inspired from percolation theory \cite{kauffman}. Sorkin and 
collaborators have conjectured that the causal set 
structure alone may be sufficient to construct a quantum theory 
of spacetime \cite{bombelli}. The poset is a discrete approximation  
of a physical manifold, which reproduces some topological 
properties of the manifold being approximated that other models are
not able to reproduce \cite{balachandran}. It has 
the structure of a topological space, with a topology defined by the 
causal order. A family of Posets of finite and increasing numbers of
points, determines a family of projective finitary topological spaces 
whose inductive limit is the continuous manifold being approximated 
\cite{sorkin91}.
Moreover, the Poset constitutes a genuine ``nonconmutative'' space from the
point of view of a generalization of Gel'fand-Naimark theorem. In effect,
to the Poset corresponds a ``nonconmutative'' $C^*$-algebra of operator
valued functions, which will be useful for constructing quantum physics
on the Poset \cite{balachandran}.

Independently of the particular mathematical construction that may
give rise to one or another choice of partial ordering, it 
seems well worth the effort to find out what can be learned 
from the structure of Causal Sets {\it per se}, insofar 
as analyzing its potential to provide a discrete 
representation of spacetime geometry and the use that this may
have in understanding various novel approaches to quantum gravity.

Progress towards what might be called 
a Causal Set representation of Quantum Gravity has been hindered by 
several factors, not least of which is the absence of a satisfactory
dynamical formulation. 

First of all, what does one mean by ``dynamical formulation'' , 
when the variables in question are the causal structure of spacetime 
itself? As often, one finds it helpful to first
answer the analogous question in Classical Mechanics. A classical 
dynamical problem can be formulated as that of finding a projection 
operator from the set of all histories onto the subset of such
histories that are solutions of the classical equations of motion. As
long as there is a single classical history corresponding to any 
given initial data set, this formulation of the dynamical problem is 
equivalent to the conventional one in terms of deterministic 
evolution equations. In quantum mechanics it is not very
meaningful to consider a single history. Instead, one would 
like to recast the dynamical problem in terms of subsets of 
the set of histories, by asking whether a subset of histories, which
is determined by particular properties, is more likely to be realised 
than its complement. 
One knows that a probability cannot be assigned to sets of histories, 
because interference leads to violations of the probability sum rules 
\cite{feynman}. Nevertheless, a meaningful interpretation can be derived 
from a {\it quantum measure} on sets of histories \cite{sorkin96}, the 
quantum measure being a generalisation of the probabilistic measure 
which takes into account the possibility of interference. We will 
adopt this point of view here, summarising it briefly in section 2.

In the case of Causal Sets, the challenge is to find a dynamical 
formulation which might explain how causal sets with 
asymptotic properties resembling those of the spacetime we live in 
might come to be selected as being at least reasonably likely.

Markopoulou and Smolin have recently proposed a dyamical Causal 
Set model where spacelike slices are spin networks which connect
to each other by means of null struts \cite{markopoulou}. 

However this construction, as any local network-building algorithm, 
suffers from a lack of Lorentz invariance at least at small scales,
due to its reliance on horizontal slices and a local scaffolding 
procedure. Whether or not effective Lorentz invariance can be
recovered at large scales, one would like to avoid introducing 
a global rest frame at the Planck scale, where the foundations 
of the theory are being set.

What is meant by the term ``Lorentz invariance'' in the present 
context? This term refers to the amplitude function (or probability) 
on the set of posets, but it is only meaningful when considering the 
amplitude of posets that can be (aproximately) embedded 
in Minkowsky space. For these posets, one may consider how the 
causal links would look in different reference frames. 
A link which in one reference frame looks to be purely timelike and
of small size, will in a highly boosted frame appear to be stretched out 
and almost null \cite{sorkin90}. A Lorentz-invariant model 
should not privilege any one reference frame over another, so 
in any given frame one should observe both short links and elongated links. 
In contrast, a local lattice-building model which in a 
given frame is only allowed to connect nearby points on a 
regular lattice, would not be a Lorentz-invariant model. 

Our purpose in this article is to propose a simple toy model with which
we will derive a quantum measure on the set of Posets 
without introducing any a priori lattice structure. We also wish 
to explore what sort of questions such a dynamical causal set model 
should be able to answer. 

In order to arrive at a model that is simple enough that 
computer computations can be performed on relatively large 
posets, we choose to set aside some of the other issues that 
have previously frustrated attempts to construct a realistic
quantum measure model for Causal Set dynamics. 
In particular, the model which we present in this article 
introduces a labeling of the points by integers, and the amplitude
is not required to be labeling invariant. We further simplify the 
problem by considering non-cyclic oriented graphs rather than posets,
the difference being that transitive relations are relevant in a 
graph whereas they are not relevant in the partial ordering.
Both invariances, labeling and transitivity, can be recovered in the 
end by summing over labelings 
and summing over all graphs that represent the same Poset. 

In section 3 we will give the outline of our toy model and present 
the corresponding quantum measure. A method to derive computable 
expressions for the measure is then described in section 4, which 
will be  applicable to sets of histories whose properties can be 
expressed by columns of the connectivity matrix. A few examples are 
evaluated numerically to reveal some of the structure of the model, 
including the measure of all histories with no black holes.  

\section{Quantum measure theory}

Quantum mechanics can be described as a simple generalization 
of classical measure theory (or probability theory). A classical measure 
is a map from an algebra of ``measurable sets'' to the positive real numbers 
which satisfies

$$I_2(A, B) \equiv \vert A \sqcup B\vert - \vert A\vert - \vert B\vert = 0,
 \eqno(2.1)$$
where $\sqcup$ denotes the union of disjoint sets. The 
``no-interference'' condition (2.1) permits a probabilistic interpretation 
for sets of histories in statistical mechanics. 

In quantum mechanics, the quantity $I_2(A, B)$ represents the 
interference term between the two sets of alternatives $A, B$,
when interference occurs the condition (2.1) is violated and
for that reason one cannot assign a probabilistic interpretation
to the sum over histories formulation. Instead of (2.1), quantum 
theory respects a slightly weaker set of conditions, which defines a 
structure known as a ``quantum measure''.

A quantum measure is positive real valued function which satisfies 
the conditions
$$\vert N\vert = 0 \Rightarrow \vert A \sqcup N\vert = \vert A\vert, 
\eqno(2.2)$$
$$I_3(A,B,C) \equiv \vert A \sqcup B \sqcup C\vert - \vert A \sqcup B\vert
- \vert A \sqcup C\vert - \vert B \sqcup C\vert + \vert A\vert + 
\vert B\vert + \vert C\vert = 0 . \eqno(2.3)$$
It is worth noting that the first axiom (2.2),
which is not necessary in probability theory because it follows from
(2.1), must be included as a separate axiom for the quantum measure
because $I_3 = 0$ in itself does not guarantee that sets with zero 
measure do not interfere with others.

Clearly, the axiomatic structure of any theory has much to say as to 
how a theory should be interpreted. Since the sum over histories 
formulation leads to a weaker structure than (2.1), one naturally
expects that quantum theory will have a weaker predictive power 
than probability theory insofar as its ability to discern which 
histories are prefered by Nature. The precise nature of this weaker
predictive power, and the correct interpretation of the sum over 
histories formulation of quantum mechanics, are encoded in the 
structure of the axioms (2.2 - 2.3). Sorkin has shown that this 
structure sustains an interpretation based on so-called ``preclusion
rules'', which establish when it can be said that a certain set of
histories is almost certain {\it not} to be realised in Nature. As one 
might expect from the form of the condition $I_3(A, B, C) = 0$, 
these preclusion rules invoke correlations between {\it three}  
events, pertaining to three disjoint regions of spacetime \cite{sorkin96}. 

It is well worth stressing that our choice to use the quantum measure 
formalism rather than, say, canonical quantization, is forced 
upon us by the absence of any {\it a-priori} causal structure.
Other pregeometrical theories, such as String Theory, may yet allow a 
canonical approach to a fundamental theory of Physics, by introducing 
a Newtonian time in an abstract world which generates our physical 
spacetime indirectly, perhaps through something like the Holographic 
Principle \cite{susskind}. However, issues regarding black hole thermodynamics 
have been raised that would eventually have to be addressed \cite{sudarsky}. 
In any event this approach is not available here: in the Causal Set 
formalism a canonical quantization would equate the abstract Newtonian 
time with the labeling of the points of the causal sets, thereby 
defeating the purpose of a pre-geometrical theory of Quantum Gravity.

In the remainder of this article we will limit ourselves to a yet weaker 
form of predictive statements than preclusion, which Sorkin refers to as 
``propensity'': When refering to a particular physical property, 
one partitions the space of histories in two disjoint subsets by
distinguishing histories which do or do not have this property.
If the measure of the set of histories which {\it do} have the property 
is much larger than its complement it can be said that it has
a high propensity. The concept of propensity is useful when analysing
the classical limit of a quantum theory. For example one might 
distinguish spacetimes that have black holes and those that do not; 
if the property of having one or more black holes has a very high 
propensity, this would constitute a prediction of the theory in the 
classical limit.

\section{A quantum measure model for directed non-cyclic graphs}

\noindent 3.1 {\it Posets, Causal sets and Directed Non-cyclic graphs}

\

As mentioned in the introduction a poset ${\cal P}$ is a discrete 
set with an antisymmetric transitive relation.

The transitivity rule (1.1) allows one to differentiate two types of
relations: the {\it links}, which are relations that cannot be obtained from 
the transitive rule, and the {\it transitive} or {\it redundant} 
relations. 

The causal structure does not depend on whether a particular relation is a 
link or a transitive relation, so in terms of pure gravity one
can say that the two types of relation are physically equivalent. 
However there is a practical difference, which shows up when performing
actual calculations or numerical simulations with causal sets. 
There is generally an enormous number of possible transitive routes 
between two points in a large Causal Set, so any algorithm which 
considers each possible route individually is only applicable to 
small causal sets, the limit being about 10 points. 
To our present knowledge, there is no generally applicable 
approximation scheme to do calculations with large causal sets.

The difficulty resides in the absence of a convenient (one-to-one)
representation of posets. The most natural approach would be to represent
a poset in terms of its ``relations matrix'', where $R_{ij} = 1$ if and
only if $x_j \preceq  x_i$ and otherwise $R_{ij} = 0$ ($x_i, x_j \in 
{\cal P}; i,j \in \IN$). The matrices {\bf R} satisfy 
the transitivity condition
$$R_{ij}=\theta(\sum_k R_{ik} R_{kj} ), \eqno(3.1)$$
where $\theta(x) = 1$ if $x > 0$ and zero otherwise. The computational 
complexity of checking these equations for each of the $N^2$ possible binary 
matrices grows like $N^5$ (using the most straightforward algorithm). 
One could equally well choose to use the {\it link matrix}, ${\bf L}$, 
where $L_{ij} = 1$ if and only if $x_i \preceq x_j$ is a link, but of course
this leads to the same computational problem. The limitation on the
number of points with which one can work affects almost
every poset-related calculation. For example, the counting of posets with a 
given number of points has only been solved up
to $N = 11$ \cite{culberson}. For large values of $N$ it has been 
shown to grow asymptotically  like \cite{daughton}  
$$C \times 2^{{N^2 \over 4} + {3 \over 2} N} e^N N^{-(N+1)},$$
but the method that yields this result does not readily generalize
to other poset calculations.

To avoid these problems, which originate from the transitivity condition,
we will consider the set of {\it all} lower-triangular
binary matrices, regardless of whether or not they include all possible
transitive relations. We will refer to these matrices as 
{\it connectivity matrices}, and denote them by {\bf C}. 

A connectivity matrix represents a directed non-cyclic graph, i.e. 
a set of points connected by arrows such that arrows do not form
closed loops. Given such a graph, it is always possible to label 
the points with consecutive integers in such a way that arrows point 
from a lesser label to a greater one. One then arrives at a 
lower-triangular binary matrix, where each entry $C_{ij}$ is equal to 
one if and only if there is an arrow in the graph from $j$ to $i$ 
(with $j < i$), and zero otherwise. Some of the connections represented
in the matrix ${\bf C}$ may be links, while others will be transitive
relations, but unlike the matrix ${\bf R}$ it is not required 
that {\it all} transitive relations be represented as connections 
or arrows of the graph. 

We will understand the connectivity matrix to represent a ``history'' 
or a posibility for ``spacetime''. The sum over histories then takes the
form of an unconstrained sum over binary arrays, which makes it relatively
easier to apply standard analytical tools. 

Of course in the end the purpose is not to compute the sum over 
all histories, but over specified subsets of histories, which 
satisfy one or another physical property of interest. A ``physical property'' 
is a property of the causal structure, i.e. one that does not pertain
either to the labeling of the points or to transitive relations.
The sum over graphs will then include the sum over all the causal orders
with the given property.

There are interesting physical properties that condition the 
connectivity matrix without increasing the computational complexity to
the same extent as condition (3.1). For such properties, a graph-based 
model can be expected to yield computable expressions. We will see some
examples in section 4.

\

\noindent {\it 3.2 The ``final question''}

\

A quantum measure model can be constructed from two basic elements: an
amplitude function on the set of histories, and a ``final
question'', $Q_f$, which by definition must refer only to the part of the 
histories in the causal future of the region of interest. The question $Q_f$
must be well-posed, in the sense that each history, $\gamma$, in the
Hilbert space, ${\cal H}$, should give one and only one answer. The answer 
set then gives a 
partition of the Hilbert space in a disjoint union of sets $E_i$, 
such that 

$$\gamma \in E_i \Leftrightarrow Q_f(\gamma) = a_i, \eqno(3.2)$$
where $a_i, i = 1, \cdots, n (n \geq 2)$, represents an element of the 
answer set.

Given an amplitude $a : {\cal H} \to \IC$, one can then construct the
following function $\vert \cdot \vert$ on the power set of ${\cal H}$
(When the computational procedure at hand does not give a finite 
answer for {\it every} subset of histories, one requires that the quantum 
measure be well-defined on a sigma-algebra of ``measureable sets''):

$$\vert A\vert = \sum_i \vert \sum_{\gamma \in A \cap E_i} a(\gamma) \vert ^2 .
\eqno(3.3)$$
One easily verifies that $\vert \cdot \vert$ satisfies the axioms (2.2-2.3) 
of the quantum measure.

We have introduced a ``final question'' as part of the procedure to
construct a quantum measure, but this question in itself is not of
any particular interest. Eventually, one would like
to be able to show that the dependence of the quantum measure on the
final question vanishes asymptotically in the limit of very large 
posets. In other words, the final question would then be an artifice 
introduced for the sole purpose of performing the 
calculation. The aim of the quantum measure formalism 
is to address other questions, which one might call  ``physical''
questions. These should be formulated in such a way that
they refer to the history {\it before} the final conditions. Physical
questions can be either about the ``present'' state of the system,
with an appropriate definition of ``present state'', or about 
the spacetime history. For each possible answer there is a set $A_i$ 
of histories with answer $a_i$, and the relative values of the 
quantum measures for different possible answers will reveal what 
the model has to say regarding this question. 

The quantum measure formalism can sometimes {\it contain} a canonically
quantized model, in the following sense. One first defines a 
one-parameter family of questions, which are to 
form a complete set, in the sense
that their combined answers describe a history completely. 
These questions can be stated as : {\it ``what is the state 
of the system at time $t$?''}. The answer set is given by the eigenstates 
of a complete set of commuting (configuration-space) observables. When 
the sets of histories corresponding to different answers $E_j^t$ do 
not interfere, {\it and} the sum of the measures over all possible answers 
at time $t$ is equal to one, then the quantum measure formalism 
will provide the same information as a canonical theory for that 
particular family of questions. This will occur when the unitarity 
condition $I_2(E_j^t, E_k^t) = 2 \delta_{jk}$ is satisfied, where
$$I_2(E_j^t, E_k^t) = \sum_i \left( \sum_{\gamma_1 \in E_j^t \cap E_i} 
\sum_{\gamma_2 \in E_k^t \cap E_i} (a(\gamma_1) a^{\ast}(\gamma_2)
+ a^{\ast}(\gamma_1) a(\gamma_2)) \right). \eqno(3.4)$$
The choice of a particular one-parameter family of questions is analogous
to a choice of ``slicing'' in canonical quantization. Other possible 
slicings, based on different choices of one-parameter families of 
questions, may well lead to different unitarity conditions. 
This sort of exercise of course is only relevant to the extent that 
one is interested in making contact with canonical quantization methods.
Taking a pre-geometrical perspective, one might  argue that 
other criteria should take precedence over unitarity in guiding the
search of the correct quantum measure; if in the end it turned out 
that such criteria were to lead one to a unique quantum measure, 
then the reverse problem of finding the
choice(s) of slicing for which a unitary canonical theory can be 
deduced would provide a satisfactory solution to the problem of time.

Here we will choose the following ``final question'': 

\

\centerline{\it ``which of the points $i < N$ emit an arrow towards $N$?''} 

\

By definition of the term ``final question'' we are assuming that 
the predictive power of this model is limited to points that do not lie 
to the future of $N$. We will label the points in such a way that $N$ 
is the largest
label and there are $N-1$ other points which may or may not be to
the past of $N$ but will certainly not be to its future.

The answer set of this question is then the set of binary words 
of length $N-1$, where a bit is equal to 0 in the absence of an arrow and 
1 denotes the presence of a connection. In terms of the connection 
matrix, the answer to the final question will be given by its last 
row, ${\Cn}_N$.

This final question generates a partition of the space of histories 
into disjoint subsets of connectivity matrices with a fixed lowest row.

There is a natural one-parameter family of questions associated to this 
particular choice of final question, namely those whose answers are the
rows of the connectivity matrix. These are the questions: ``which points 
emit arrows towards the point labeled by $t$?''. Note that in this case
the cardinality of the answer set grows like $2^{t-1}$.

\

\noindent {\it 3.3 The amplitude and the quantum measure}

\

We will make the following ansatz for the amplitude (factorizability):
$$a({\bf C})\eqdef \prod_{m=2}^{N} a({\Cn}_{m-1}, m-1 \to {\Cn}_{m}, m), 
\eqno(3.5)$$
where
$$a({\Cn}_{m-1}, m-1 \to {\Cn}_m, m) \eqdef A_{C_{m m-1}} \prod_{l < m-1} 
A_{C_{ml} C_{m-1 l}}. \eqno(3.6)$$

We will also require that the amplitude to create a connection  ${\Cn}_m$ 
be independent of the previous state, $\Cn_{m-1}$. Choosing 
$$A_0 = A_{00} = i A_{01} = \sqrt{q}, \eqno(3.7)$$
$$A_1 = A_{11} = i A_{10} = \sqrt{p} \eqno(3.8)$$
and $p+q = 1$ ($p, q \in \IR_{>0}$), we arrive at a quantum generalisation 
of a one-dimensional directed percolation model. In the one-dimensional
directed percolation model, $N$ points are labeled by consecutive integers 
as in our model, and each point can connect to any of the previous points
with a constant probability $p$. 

The propagator (3.6) then becomes
$$a({\Cn}_{m-1}, m-1 \to {\Cn}_m, m) = (\sqrt{q})^{m-1-\sum_l C_{ml}} \  
(\sqrt{p})^{\sum_l C_{ml}} \ 
(-i)^{\sum_i (1 - \delta_{C_{m i} C_{m-1 i}})}. \eqno(3.9)$$
Using (3.4) and substituing (3.5), one finds that the model would be 
unitary with the slicing $\lbrace {\Cn}_m; m=1,2,...\rbrace$ if one
chose 
$$\sum_{{\Cn}_m} a({\Cn}_{m-1}, m-1 \to {\Cn}_m, m)a^{\ast}({\Cn}_{m-1}, m-1 \to 
{\Cn}_m, m) = \prod_{l=1}^{m-1}  \delta_{C_{m-1l} C^{\prime}_{m-1l}},$$
i.e. if $p=q=1/2$. 

To simplify the notation, we will introduce the total number 
of entries equal to one in the $m$'th column of the connectivity matrix, 
$$C_m = \sum_{i = 1}^{N} C_{im} , \eqno(3.10)$$
and the number of ``kinks'' in each column,
$$K_m = \sum_{i = m+1}^{N-1} (1 - \delta_{C_{im} C_{i+1 m}}) . \eqno(3.11)$$
The sum of these quantities over all of the columns of the connectivity 
matrix yield the total connectivity $C$ and total kink number $K$, 
respectively.

We then arrive at a simple expression for the amplitude of a 
connectivity matrix (or ``history''),
$$a({\bf C}) = (\sqrt{p})^C (\sqrt{q})^{C^N_2 - C} (-i)^K, 
\eqno(3.12)$$
where $C^N_2$ is the binomial coefficient.

The quantum measure of a set $A$ of connectivity matrices is then given$$\vert A \vert = \sum_{{\Cn}_N} \vert \psi(A, {\Cn}_N, N) \vert ^2,
\eqno(3.13)$$
where 
$$\psi(A, {\Cn}_N, N) \eqdef \sum_{{\bf C} \in A: {\Cn}_N fixed} a({\bf C}). 
\eqno(3.14)$$

\section{Analytical and Numerical Results}

\noindent {\it 4.1 Measure of the space of histories}, $\vert {\cal H} \vert$

\

As a first example one can compute the measure of the space ${\cal H}$
of all possible non-cyclic oriented graphs of $N$ points. In that case,
one can drop the label $A$ in (3.14) and write
$$\vert {\cal H} \vert = \sum_{{\Cn}_N} \vert \psi ({\Cn}_N, N)\vert ^2. 
\eqno(4.1)$$
The argument of the square modulus can be loosely interpreted as a 
``cosmological wave function''. To compute
$$\psi({\Cn}_N, N) \eqdef \sum_{{\bf C}: {\Cn}_N fixed} 
(\sqrt{p})^C (\sqrt{q})^{{C^N_2} - C} (-i)^K, \eqno(4.2)$$
we use the fact that $C = \sum_m C_m$ and $K = \sum_m K_m$ to
factorize the expression above and consider each column of the 
connectivity matrix individually:
$$\psi({\Cn}_N, N) = (\sqrt{q})^{C^N_2} 
\prod_{m=1}^{N-1} \psi_m (C_{Nm}, N), 
\eqno(4.3)$$
$$\psi_m (C_{Nm}, N)\eqdef \sum_{C_m=C_{Nm}}^{N-m+C_{Nm}-1} 
\sum_{K_m(C_m)} (\sqrt{p \over q})^{C_m} 
(-i)^{K_m} N_{C_{Nm}}(K_m, C_m), \eqno(4.4)$$
where $N_{C_{Nm}}(K_m, C_m)$ is the number of  binary words 
of $N-m-1$ bits with $C_m$ bits equal to 1 and $K_m$ kinks, when
the last bit in the column has been set equal to $C_{Nm}$.

The functions $N_{C_{Nm}}(K_m,C_m)$ can be calculated by considering 
the number of ways of making $k$ cuts in a sequence of $C_m$ ones 
and inserting the zeroes at the cuts. One finds
$$N_0(2k, C_m) = \delta(k, 0) \delta(C_m,0) + \pmatrix{C_m - 1 \cr k-1}
\pmatrix{N-m-1-C_m \cr k}, \eqno(4.5)$$
$$N_1(2k, C_m) = \delta(k, 0) \delta(C_m, N-m) + \pmatrix{C_m - 1 \cr k}
\pmatrix{N-m-1-C_m \cr k-1}, \eqno(4.6)$$
$$N_0(2k+1, C_m) = \pmatrix{C_m - 1 \cr k} 
\pmatrix{N-m-1-C_m \cr k}, \eqno(4.7)$$
$$N_1(2k+1, C_m) = \pmatrix{C_m - 1 \cr k}
\pmatrix{N-m-1-C_m \cr k}. \eqno(4.8)$$

To determine the bounds on $K_m$, we must regard its dependence on 
$C_{Nm}$ and make a comparison between the number of bits equal 
to 1, $C_{m}$, and the number of bits equal to zero, $N-m-C_m$, 
analyzing the different possible arrangements. The result is:

if $N-m-C_m < C_m \leq N-m+C_{Nm}-1 \Longrightarrow$
$$K_m \in [1-C_{Nm}, 2(N-m-C_m+C_{Nm}-1)+1-C_{Nm}],$$

if $N-m-C_m=C_m \Longrightarrow K_m \in [1,2C_m-1]$,

if $C_{Nm} \leq C_m < N-m-C_m \Longrightarrow
 K_m \in [C_{Nm},2C_m-C_{Nm}]$.

These equations allow one to calculate the function $\psi({\Cn}_N, N)$ 
on a computer and compute the measure $\vert {\cal H} \vert$. Not surprisingly,
this measure is equal to 1 in the ``unitary case'' $p = q = 1/2$ when
$\psi$ can be interpreted as a cosmological wave function. This 
is not particularly relevant in the context of the quantum measure 
interpretation.

\

\noindent {\it 4.2 Graphs with a Fixed Number of Arrows from a Given Point}

\

 Setting aside for the time being the issue of labeling invariance, 
we will compute the propensity that a point with a given label emit
a fixed number of arrows towards other points of the graph. Let
$A_{C_l}(l)$ be the set of histories (graphs) where point labeled $l$ 
emits $C_l$ outgoing arrows. From (3.12-3.14) we have

$$\vert A_{C_l}(l) \vert = \sum_{{\Cn}_N} 
\vert \psi(A_{C_l}(l), {\Cn}_N, N) \vert ^2,
\eqno(4.9)$$
where 
$$\psi(A_{C_l}(l), {\Cn}_N, N) = (\sqrt{q})^{C^N_2} 
(\sqrt{p\over q})^{C_{N N-1}}
\psi_l(A_{C_l}(l), C_{Nl}, N) \prod_{\matrix{{\s m=1} \cr 
{\s m \neq l} \cr}}^{N-2} \psi_m(C_{Nm}, N) \eqno(4.10)$$
and
$$\psi_l(A_{C_l}(l), C_{Nl}, N) \eqdef \sum_{\Vn _l : A_{C_l}(l)} 
(\sqrt{p \over q})^{C_l} (-i)^{K_l}, \eqno(4.11)$$
where we have used the short-hand notation $\Vn _l : A_{C_l}(l)$ to denote
the sum runs over the columns of binary words $\Vn _l$, which satisfy that 
the number of connections $C_l$ be fixed. One finds

$${{\vert A_{C_l}(l) \vert } \over {\vert {\cal H} \vert }} = 
({p\over q})^{C_l} {{\vert Z_0 \vert ^2 + \vert Z_1 \vert ^2 } \over
{\vert \psi_l(0, N) \vert^2 + \vert \psi_l(1, N) \vert^2} }, 
\eqno(4.12)$$
where 
$$Z_{C_{Nl}} = \sum_{K_l} (-i)^{K_l} N_{C_{Nl}}(K_l, C_l)\eqno(4.13)$$
represents the quantum interference factor. This ratio was computed 
numerically for $p = q = 1/2$, as a function of $C_l$.

\begin{figure}
\begin{center}
\epsfysize=5cm
\epsfbox{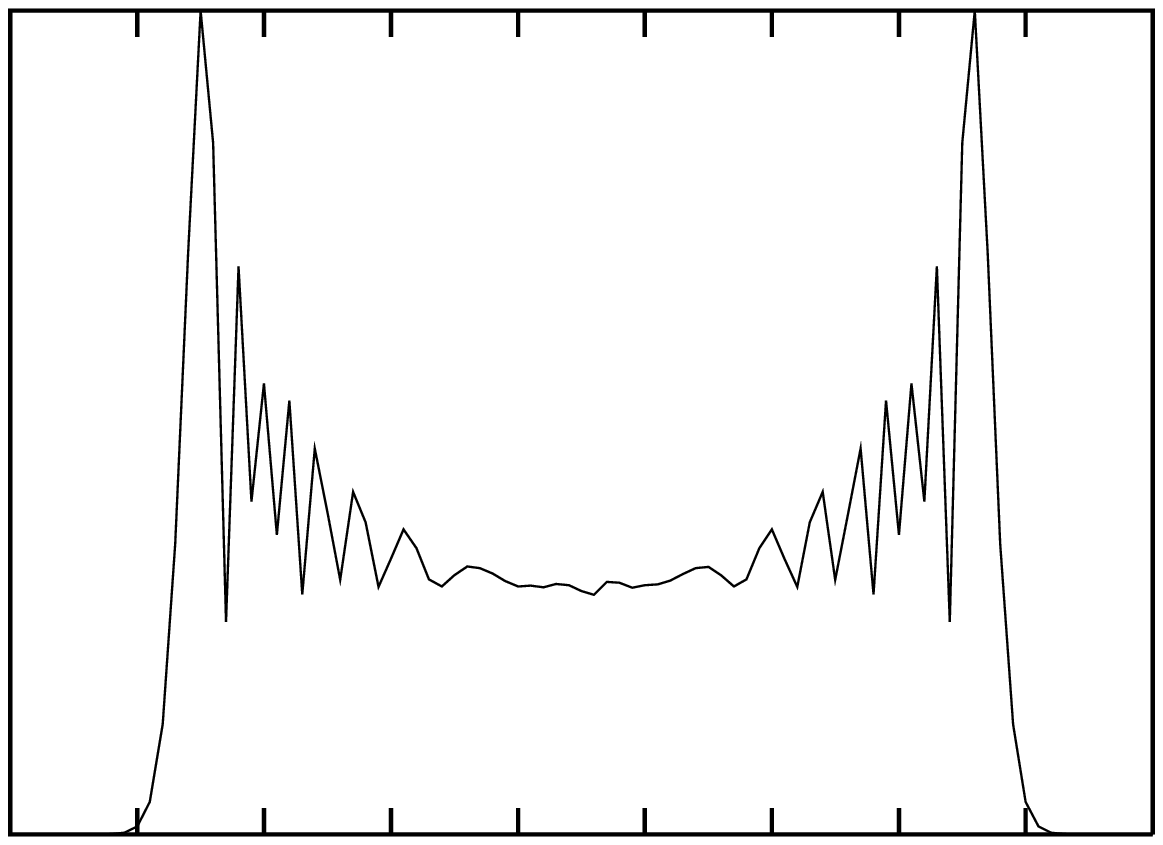}
\caption{The propensity that a point (point number 90 
from the final point) emits $C$ arrows is represented as a function of $C$,
for the case $p=q=1/2$.
The sharp rise and fall at both ends are remnants of a classical binomial 
distribution, whilst in the middle quantum interference is observed. Contrary
to the classical case the propensity does not peak at $C=45$, half the
possible number of arrows.}
\end{center}
\end{figure}

In the classical directed percolation model, the probability of $C_l$ is 
given by a binomial distribution 
$$p(C_l) = \pmatrix{N-l-1 \cr C_l} p^{C_l} (1-p)^{N-l-1-C_l}. \eqno(4.14)$$
The ratio of the quantum measure to its classical counterpart,
$$F(l) = {{\vert A_{C_l}(l) \vert} \over {p(C_l)}} \eqno(4.15)$$
is a form factor which can be interpreted as representing the effect of 
quantum interference. A comparison between [Figure 1] and the binomial 
distribution reveals that the destructive interference is most important
at the midpoint where half the available bits are equal to one.
When $C_l$ is nearer one of the two extremal values one observes the
same exponential dropoff as in the statistical model.

In the crossover between these two regimes, a striking interference 
pattern is observed. To witness the origin of this interference we show
the real part of the form factor $Z_0$ (the imaginary part is 
its mirror image) [Figure 2].

\begin{figure}
\begin{center}
\epsfysize=5cm
\epsfbox{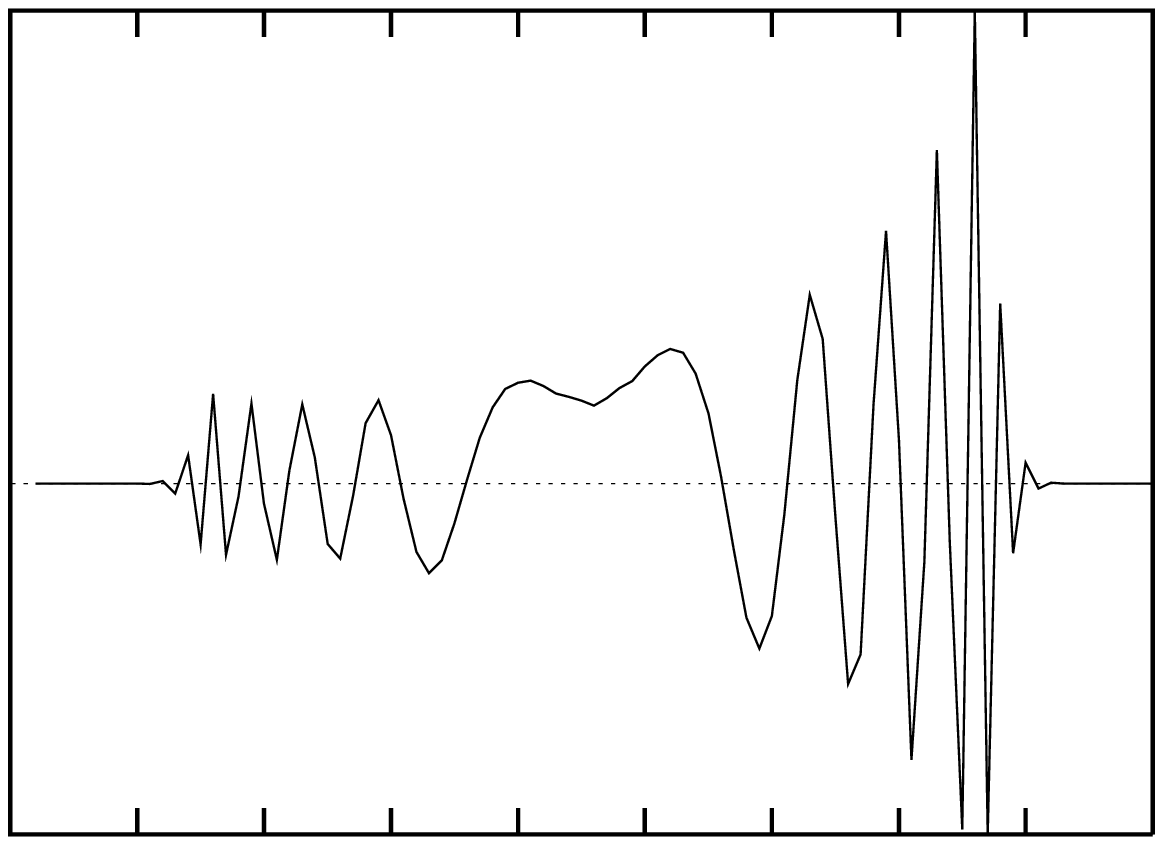}
\caption{The real part of the form factor which gave rise
to the previous figure is represented here.}
\end{center}
\end{figure}

\

\noindent {\it 4.3 Graphs with No Arrows from a Given Point}

\

When the number of out-arrows from a point is equal to zero 
this point is a sink in the oriented graph. It is then 
not emitting any outgoing signals and therefore might
loosely be called a classical ``black hole''. The propensity for
a point labeled $l$ to be a sink is given by

$${{\vert A_0(l) \vert } \over {\vert \Abar \vert }} = {{\sum_{{\Cn}_N} \vert 
\psi(A_0(l), {\Cn}_N, N) \vert ^2} \over {\sum_{{\Cn}_N} \vert 
\psi(\Abar, {\Cn}_N, N) \vert ^2}}, \eqno(4.16)$$
where $\Abar$ is the the complement of the set $A_0(l)$. 

\noindent With $A_0(l) = \{ {\bf C} : {\Vn}_l = (0,0, \cdots, C_{Nl}), 
l \notin \{ 1, N-1, N\} \}$, we have 
$$\psi(\Abar, {\Cn}_N, N) = \psi({\Cn}_N, N) - \psi(A_0(l), {\Cn}_N, N), 
\eqno(4.17)$$
$${{\vert A_0(l) \vert } \over {\vert \Abar \vert }} = { {1 + {p \over q}}
\over {\vert \psi_l(0, N) - 1 \vert ^2 + \vert \psi_l(1, N) + 
i \sqrt{p \over q} \vert ^2} }.
\eqno(4.18)$$

The sink propensity is represented as a function of the parameter $p$ in 
[Figure 3], for $N-l = 100$ and $200$. On the same figure we represented the 
corresponding classical expression $q^{N-l} / (1-q^{N-l})$ 
for comparison. Not surprisingly the classical curve crosses 
the line $y=1$ at a value of $p$ that is just a bit lower than 
the percolation point $1/N$. The ratio of the quantum measures
in contrast remains close to this line for a range of values of $p$
covering three orders of magnitude. The peak in [Figure 3b] 
indicates a high sink propensity at $N-l = 200$ for $p \approx 0.036$.
This peak shows that an effective inhomogeneity is induced from 
the dependence of the measure on the labeling of the points. 
This labeling introduces the order of the natural numbers. Natural
ordering enters in two different places: in the 
definition of a ``kink'', and in the notion of a ``sink'', which treats 
out-arrows differently from in-arrows and thereby gives relevance to the 
``distance'' $N-l$. The existence of the points between $N$ and $l$ is not
relevant for the causal order.

\begin{figure}
\begin{center}
$\matrix{ {\epsfysize=10cm \epsfbox{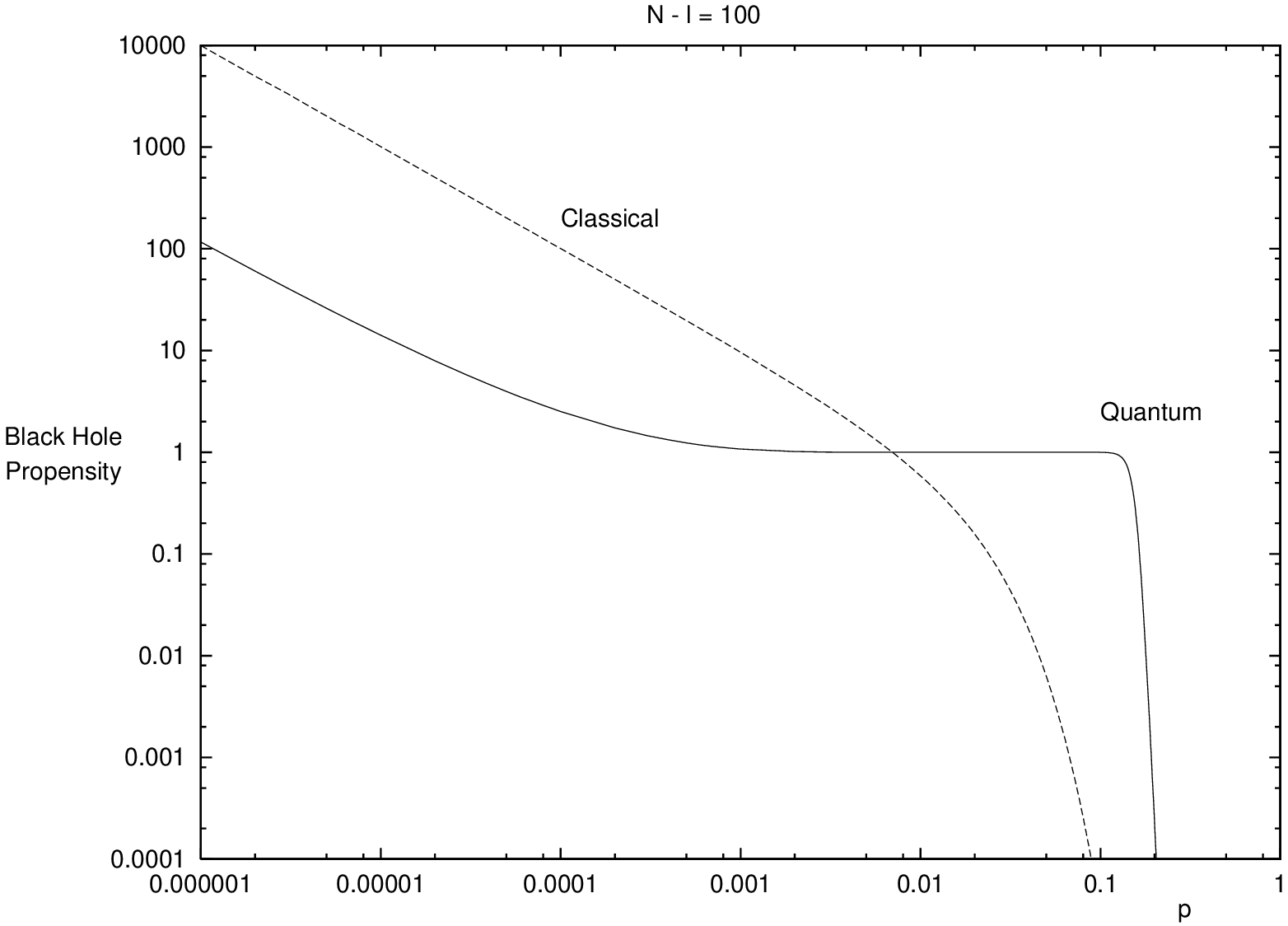}}
& {\epsfysize=10cm \epsfbox{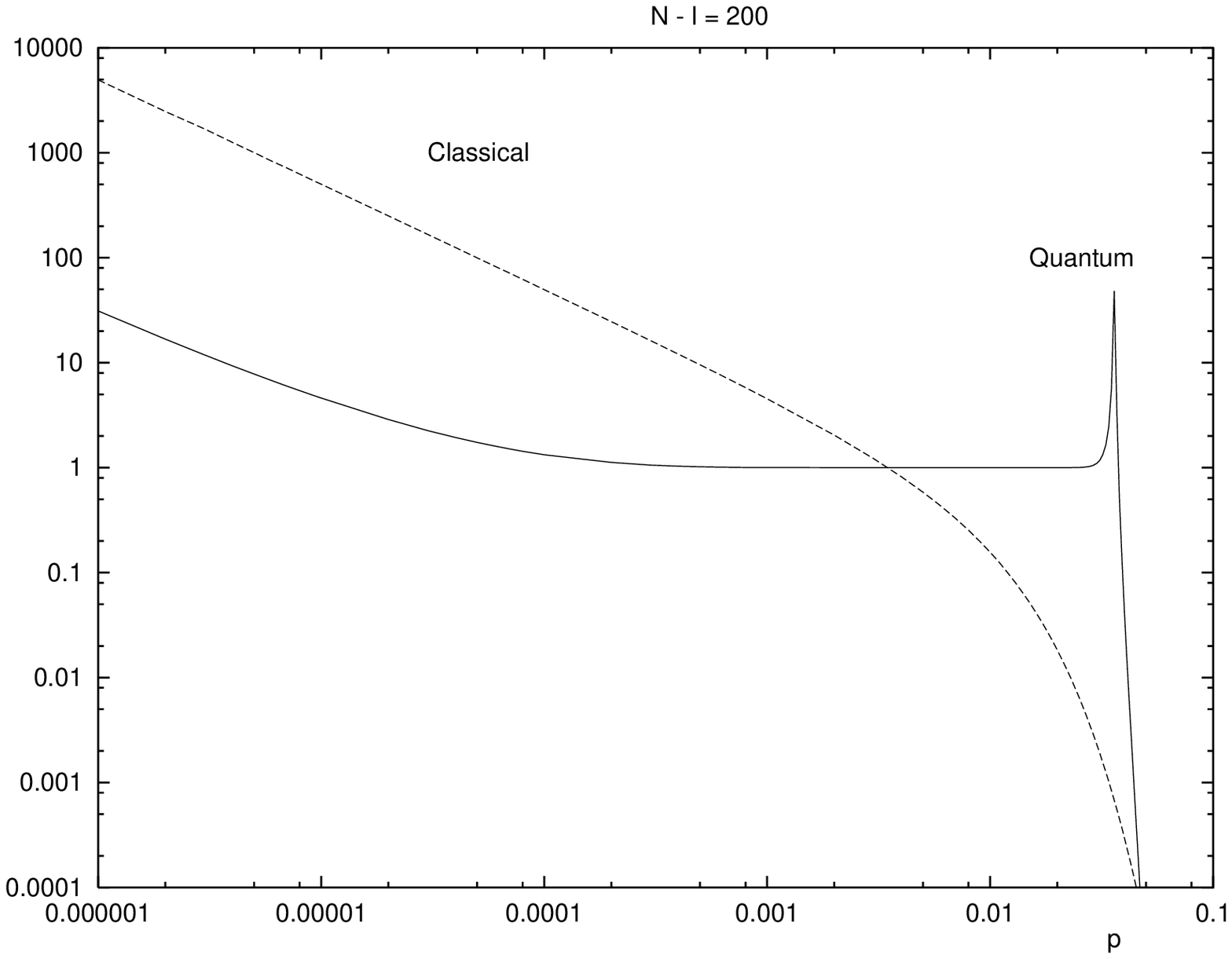}} \cr}$
\caption{For lower values of $p$ the classical and quantum 
theories behave differently: in classical percolation one has a phase
transition at $p_c \sim 1/N$ 
where the propensity that an arbitrary point is a sink is about 
equal to the propensity that it is not a sink. This is shown in the figure 
as a dotted line which represents the ratio of these two propensities. 
In the quantum model this ratio remains equal to one over an extended 
range of values of $p$. For some points such as $N-l = 200$ (Fig. 3b), 
a peak in the black hole propensity ratio can occur at large values 
of $p$, contrary to classical intuition.}
\end{center}
\end{figure}

To analyze this effective inhomogeneity, the code was modified to 
scan the p-axis for a possible peak, for every value of $N-l$.
This effort was distributed over several computers to arrive at
the results represented in [Figure 4]. Very high peaks are observed 
for particular values of $N-l$, signaling points that almost certainly 
will be sinks. 

\begin{figure}
\begin{center}
\epsfysize=12cm
\epsfbox{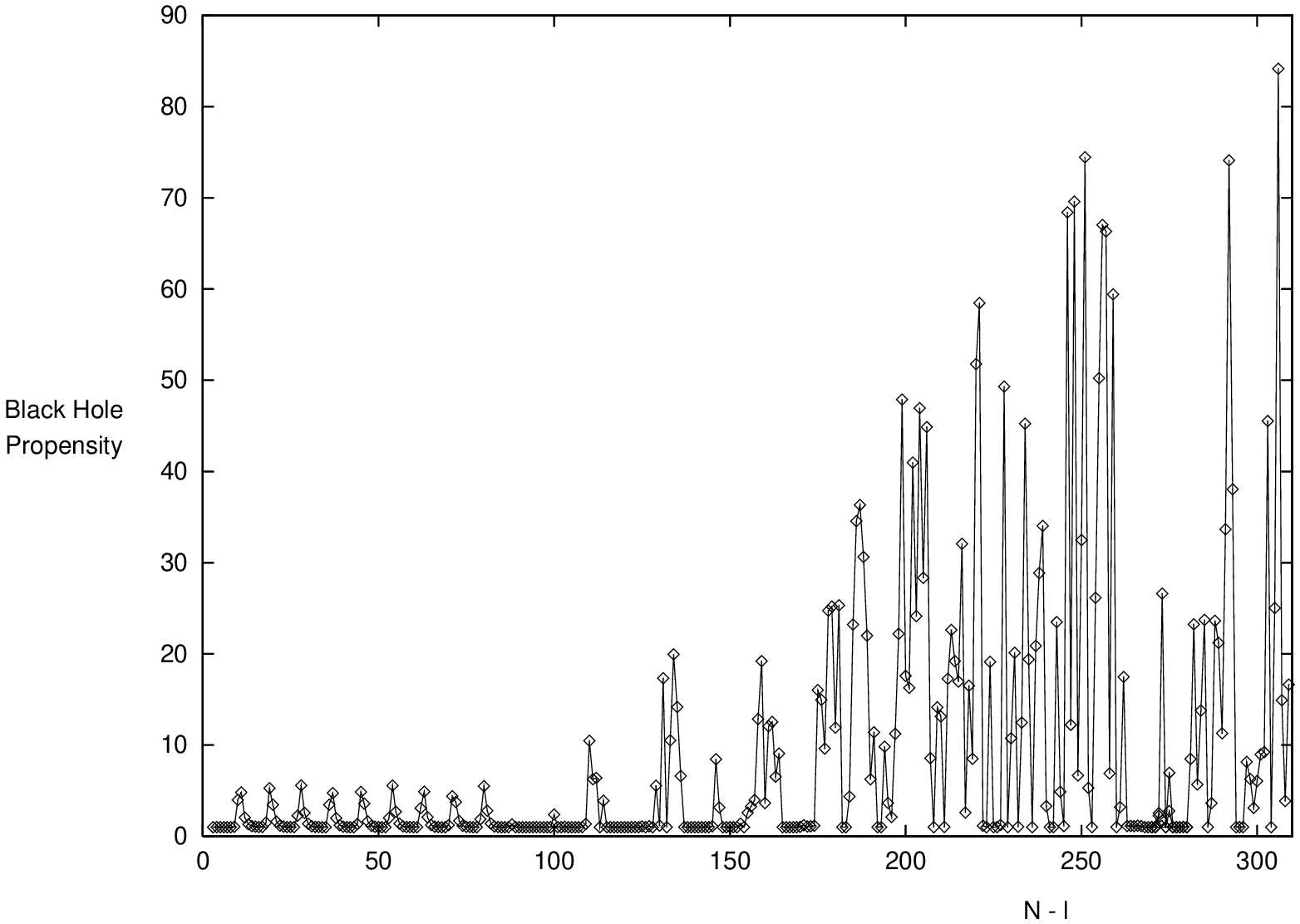}
\caption{When a peak in the black hole propensity ratio is
observed, the peak value of the propensity ratio is given in this graph.
Different points here are obtained with different values of $p$, for
each point the peak is located first and only the highest point is 
represented. Some points like $N-l = 306$ can produce a significantly 
large ratio, indicating the likely occurrence of a sink for a 
particular value of the parameter $p$.}
\end{center}
\end{figure}

\

\noindent {\it 4.4 The Measure of Spacetimes with No Black Holes}

\

An example of a labeling invariant question is ``does spacetime have 
no black holes?''. Let $\Bbar$ denote the set of graphs with
no sinks and its complement $B_0$ the set of graphs with one or more
sinks. Then,

$$\psi(\Bbar, \Cn_N, N) = (\sqrt{q})^{C^N_2} \prod_{m=1}^{N-1} \ 
\sum_{C_m, K_m} \ \left(\sqrt{p \over q}\right)^{C_m} (-i)^{K_m} 
\ N_{C_{Nm}}(K_m, C_m) \ \left(1 - \delta(C_m - C_{Nm})\right) \eqno(4.19)$$
and
$$\psi(B_0, \Cn_N, N) = \psi(\Cn_N, N) - \psi(\Bbar, \Cn_N, N). \eqno(4.20)$$
Using the above equations one finds
$${{\vert \Bbar \vert } \over {\vert B_0 \vert }} = {{\prod_m \left(
\vert \psi_m(0) - 1 \vert^2 + \vert \psi_m(1) + i \sqrt{p \over q}\vert^2 
\right)} \over {\prod_m \left( \vert \psi_m(0) \vert^2 + 
\vert \psi_m(1)\vert^2 \right) - 2 Re(W) + \prod_m \left( \vert \psi_m(0)
- 1 \vert^2 + \vert \psi_m(1) + i \sqrt{p \over q} \vert^2\right)}}, 
\eqno(4.21)$$
where
$$W = \prod_{m=1}^{N-2} \left( \vert \psi_m(0)\vert^2 - \psi_m(0) + \vert 
\psi_m(1) \vert^2 - i \sqrt{p \over q} \psi_m(1) \right). \eqno(4.22)$$

The sink propensity is very large for small values of $p$, since 
connections in that case are rare for statistical reasons, and vice
versa it will drop to zero as the connection probability becomes large. 
A numerical evaluation of (4.21-4.22) shows however that the
quantum model once again displays a broad region where the 
ratio is equal to one and the model simply does not provide any
information as to whether or not one should expect to find 
any sinks [Figure 5]. A peak 
in the sink propensity for values of $p$ well above the classical 
percolation point at $p = 1/N$ reflects the presence of quantum 
interference effects.

\begin{figure}
\begin{center}
\epsfysize=12cm
\epsfbox{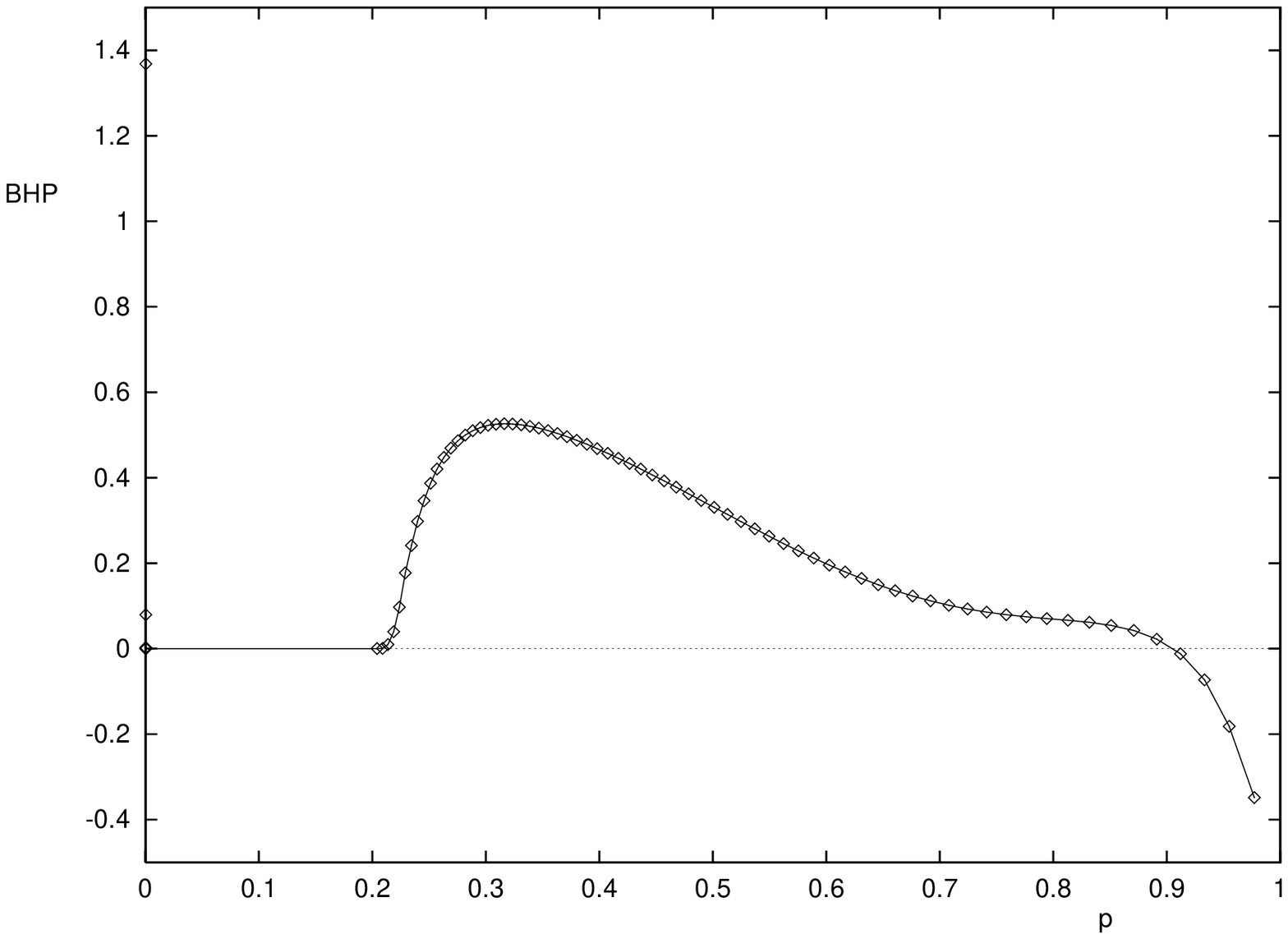}
\caption{The ``black hole propensity'' is represented as 
a function of the connectivity parameter $p$. }
\end{center}
\end{figure}

\section{Discussion}

We presented a general procedure to construct a quantum model 
in the quantum measure formalism, and applied this procedure to
arrive at  a toy model of causal set dynamics. 

Technical problems related to the transitivity condition on causal
relations were circumvented by considering
{\it non-cyclic oriented graphs} in place of posets. In a graph,
one distinguishes transitive relations that are represented with 
an arrow to those that are not, even though this distinction is not
relevant in terms of the causal order. Yet, physically relevant 
information can be derived from this model, by considering a 
limited class of physical properties that condition the 
connectivity matrix without raising the computational complexity to
the same extent as the transitivity condition on a relations matrix 
(3.1). 

An alternate strategy would be to regard the information in representing
a transitive connection as an arrow to be physically significant and
related to the propagation of a signal between two points. One would 
then be dealing with a larger theory, which describes gravity together 
with some other discrete excitations with a role similar to vector bosons. 

This model may be relevant to other areas than causal sets and quantum
gravity, if viewed as a quantum mechanical model for percolation. 
In this light, it is well worth recalling the observation that the
quantum model displays two distinct transitions, from an ``underconnected''
phase where almost all points are isolated, through a broad ``quantum
phase'', and finally to a ``connected''
phase where almost all points belong to the same dominant cluster. This
constrasts with a classical percolation model where the transition 
between the first and last phases occurs at a critical point. A 
similar situation is often found in statistical models such as 
traffic models and other self-organized critical systems. 
In a traffic model for example, the forcing parameter is the rate at
which vehicles are placed at one end of the system looking for an
opportunity to proceed. With low forcing one finds a constant
mean vehicle speed limited only by the circulating velocity and the
presence of traffic lights. At intermediate values of the forcing 
parameter the entry point can become obstructed and this 
limits further increases in the flow of traffic. This phase is
characterized by a scale-invariant behaviour of the rate of flow. At
yet higher forcing levels, a second phase transition takes place, to
a completely saturated situation with an exponential cutoff in
velocity fluctuations. The possibility that the quantum generalization 
of a directed percolation model produce phenomena similar to those that are 
generically found in self-organized critical phenomena is of course
tantalizing, particularly if this can help avoid the need to fine-tune
parameters of the system to match theory with experiment. 

The measure calculation of section (4.2) hints that it may be
possible to eliminate the need for fine-tuning a theory at the
critical point without breaking Lorentz invariance. The extended 
plateau in sink propensity shows that there is a large range of values 
of $p$ where the quantum model behaves neither like a classical 
model below the critical point nor like a classical model above
the critical point. We do not know whether this plateau indicates 
the presence of a different ``quantum percolation phase'', with 
properties unlike those of any classical stochastic model, or 
the extension of a classical critical regime characterized by a power-law 
distribution of cluster sizes, as in self-organized critical systems.

To make statements about percolation more precise one 
would have to consider the
measure of the set of histories such that two given points $k,l$ 
are  related, either by a direct connection or by a transitive
chain of connections. Unfortunately this immediately leads one
into the difficulty that we tried to avoid by working with the 
connection matrix rather than the relations matrix: the transitivity
condition requires that one consider powers of the matrix ${\bf C}$ 
up to an order that increases linearly with $N$, only to determine
whether or not there exists a transitive relation between two 
points. This implies that the measure cannot be computed using 
factorization, as in section 4, and calculation by exhaustive
summation is limited to about $N=11$ with a modern supercomputer.

\subsubsection*{Acknowledgements} 

This work was partially supported through the Spinoza grant of 
the NWO (The Netherlands), and DGAPA-UNAM grant number IN105197.
The authors wish to express their gratitude to Rafael Sorkin who 
has been one of the chief motivators of this project. We also wish 
to thank the participants of the quantum gravity
meetings at Utrecht for many enlightening discussions, particularly
Gerard 't Hooft, Serge Massar, Henk Stoof, Henk Van Beieren 
and Eric Verlinde.

\vfill\eject

\vfill\eject

\section{Figure captions}

\noindent Figure 1. The propensity that a point (point number 90 
from the final point) emits $C$ arrows is represented as a function of $C$,
for the case $p=q=1/2$.
The sharp rise and fall at both ends are remnants of a classical binomial 
distribution, whilst in the middle quantum interference is observed. Contrary
to the classical case the propensity does not peak at $C=45$, half the
possible number of arrows.

\

\noindent Figure 2. The real part of the form factor which gave rise
to the previous figure is represented.

\

\noindent Figure 3. For lower values of $p$ the classical and quantum 
theories behave differently: in classical percolation one has a phase
transition at $p_c \sim 1/N$ 
where the propensity that an arbitrary point is a sink is about 
equal to the propensity that it is not a sink. This is shown in the figure 
as a dotted line which represents the ratio of these two propensities. 
In the quantum model this ratio remains equal to one over an extended 
range of values of $p$. For some points such as $N-l = 200$ (Fig. 3b), 
a peak in the black hole propensity ratio can occur at large values 
of $p$, contrary to classical intuition.

\

\noindent Figure 4. When a peak in the black hole propensity ratio is
observed, the peak value of the propensity ratio is given in this graph.
Different points here are obtained with different values of $p$, for
each point the peak is located first and only the highest point is 
represented. Some points like $N-l = 306$ can produce a significantly 
large ratio, indicating the likely occurrence of a sink for a 
particular value of the parameter $p$.

\

\noindent Figure 5. The ``black hole propensity'' is represented as 
a function of the connectivity parameter $p$.

\vfill\eject

\begin{figure}
\begin{center}
\epsfbox{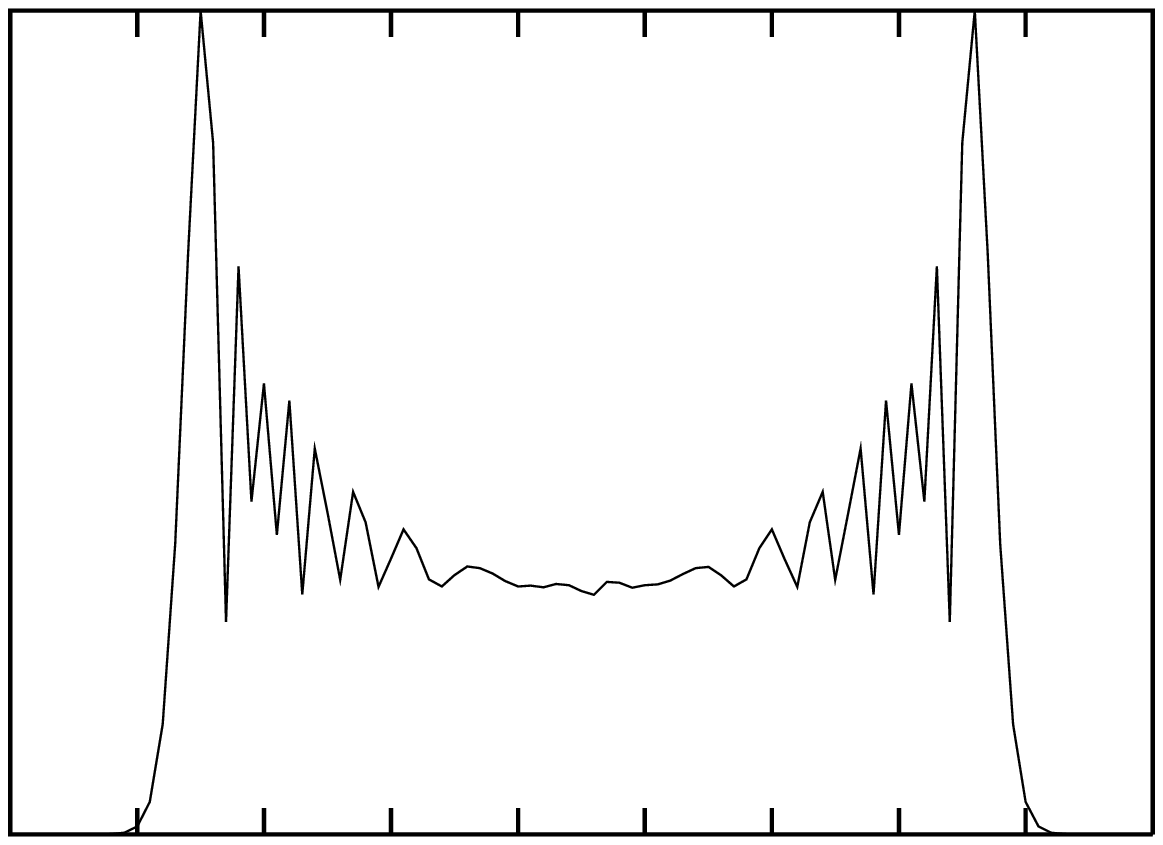}
\end{center}
\end{figure}

\begin{figure}
\begin{center}
\epsfbox{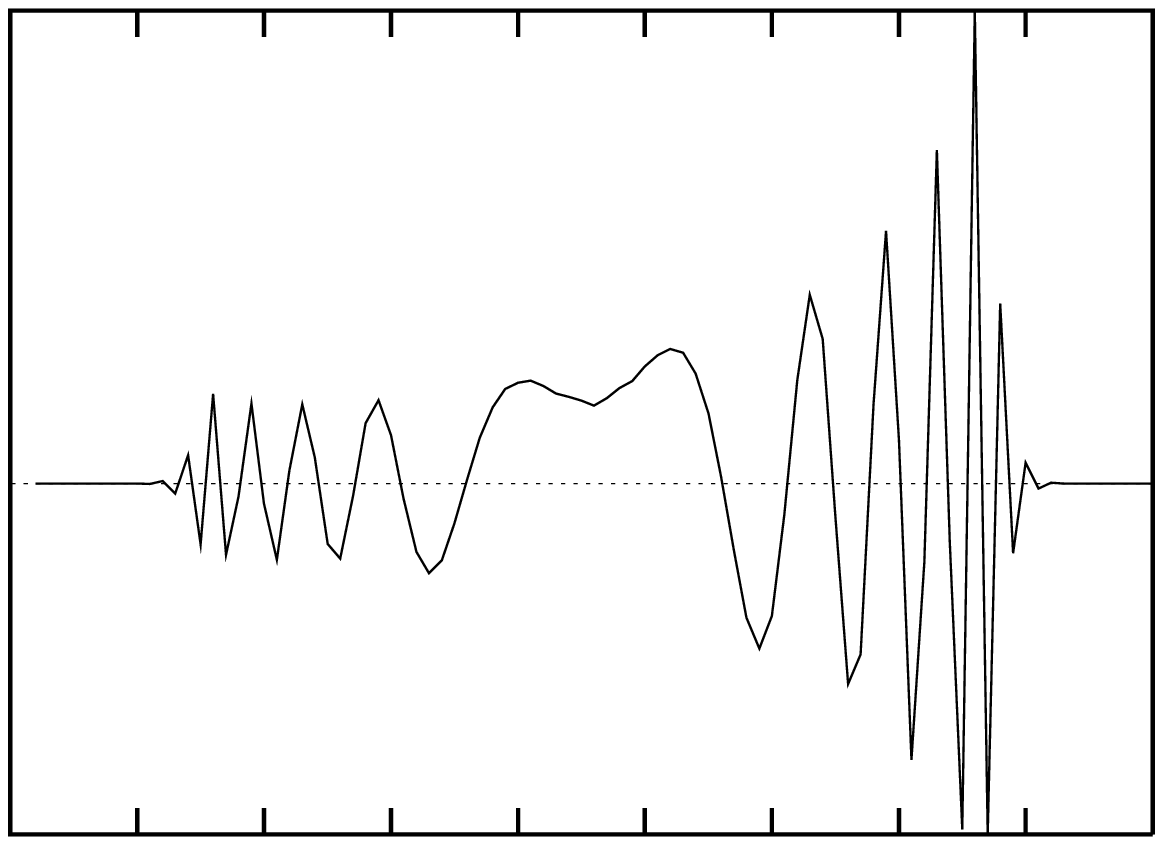}
\end{center}
\end{figure}

\begin{figure}
\begin{center}
\epsfbox{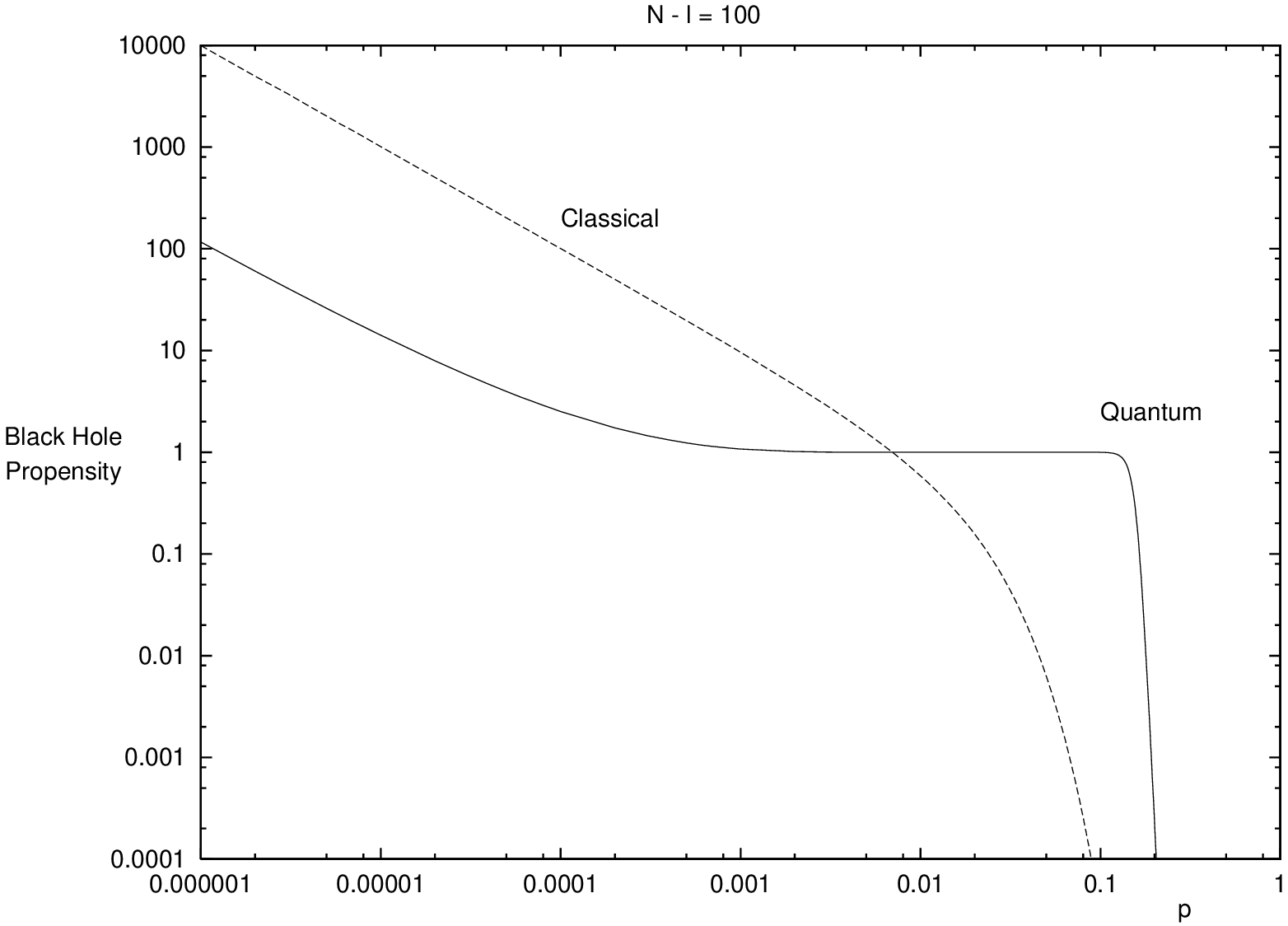}
\end{center}
\end{figure}

\begin{figure}
\begin{center}
\epsfbox{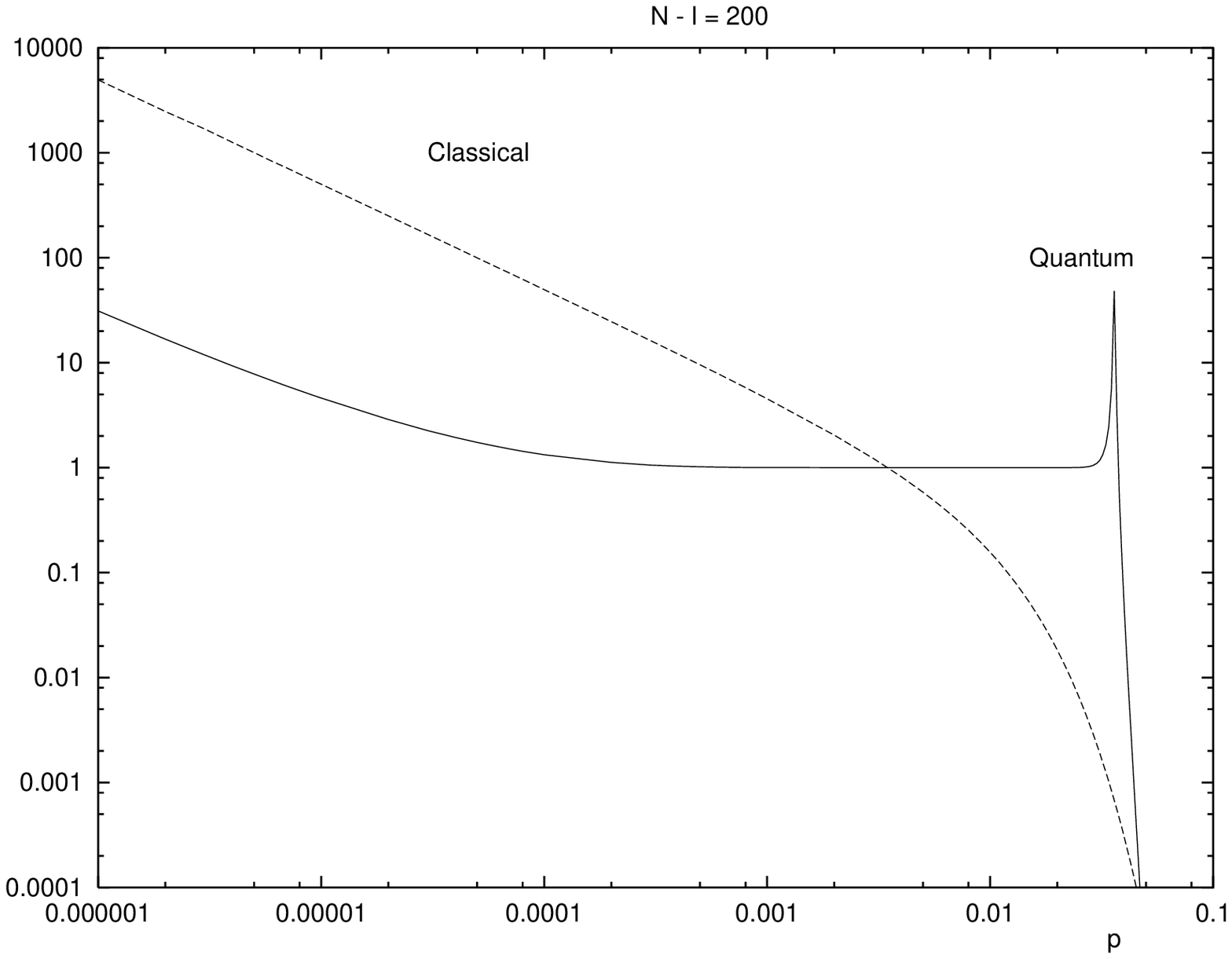}
\end{center}
\end{figure}

\begin{figure}
\begin{center}
\epsfbox{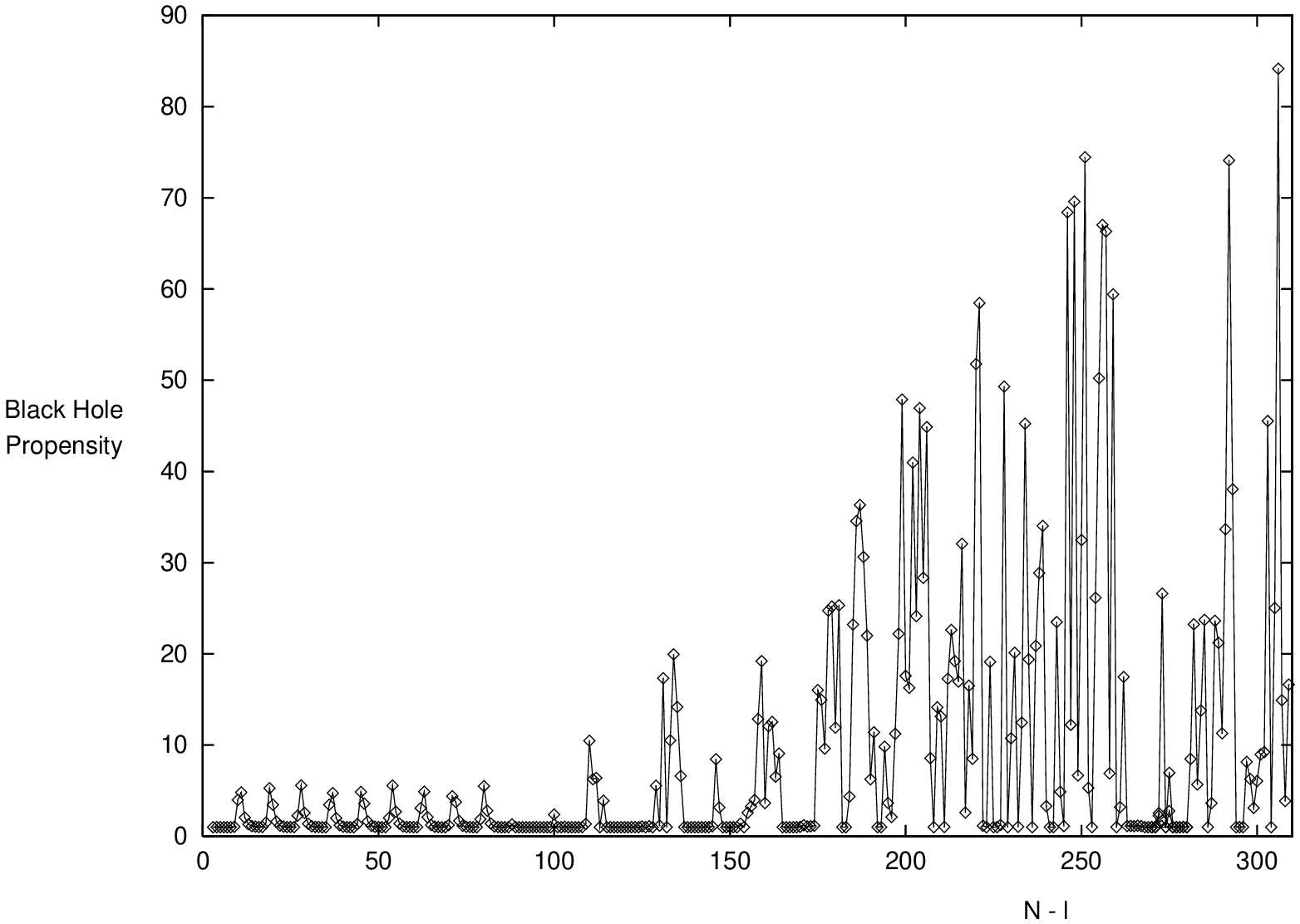}
\end{center}
\end{figure}

\begin{figure}
\begin{center}
\epsfbox{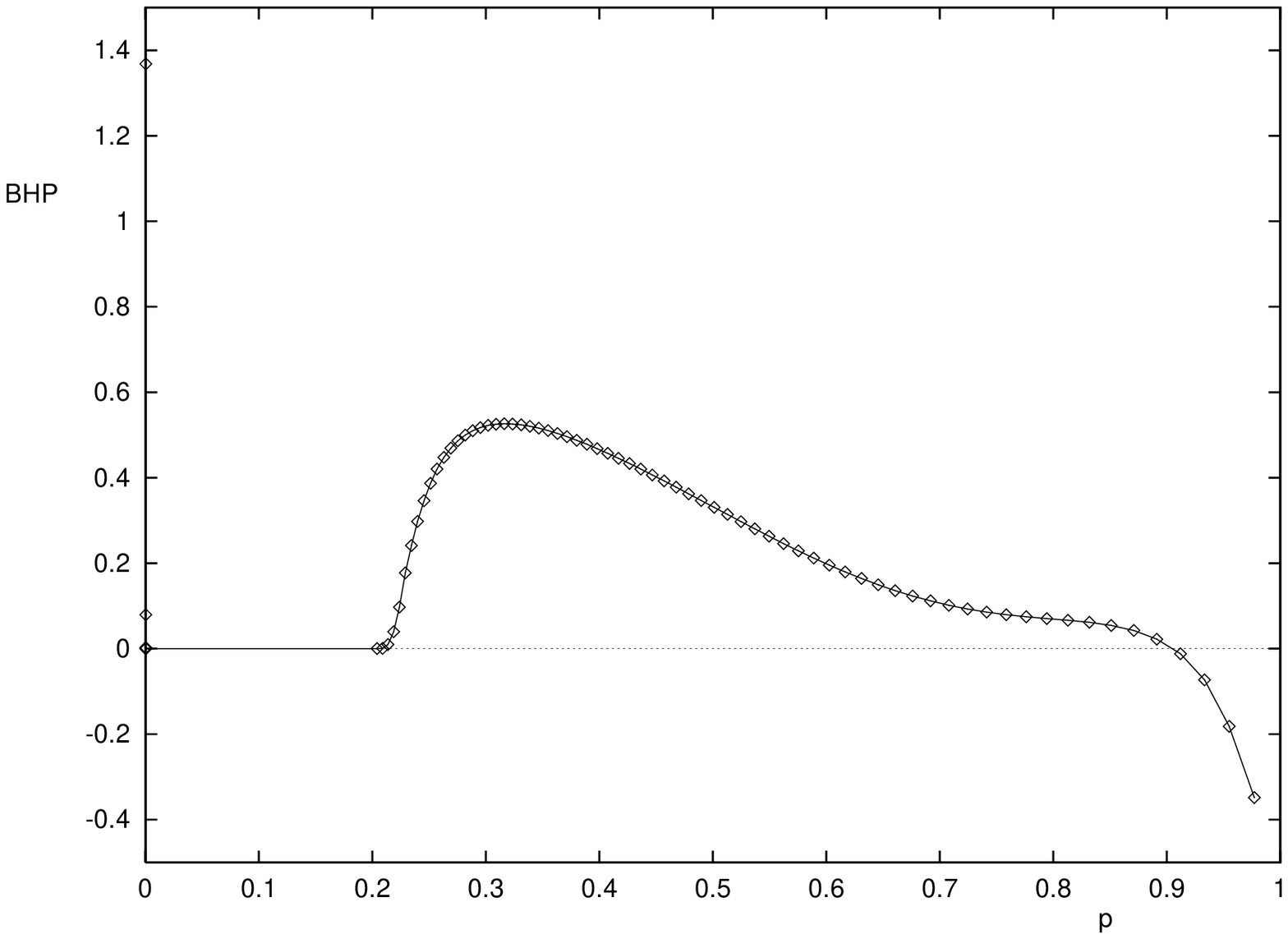}
\end{center}
\end{figure}

\end{document}